\def\ra{\rangle}
\def\la{\langle}
\def\Nb{{\Bbb N}}
\def\Rb{{\Bbb R}}

\def\be{\begin{equation}}
\def\ee{\end{equation}}
\def\ba{\begin{array}}
\def\ea{\end{array}}

\def\ra{\rangle}
\def\la{\langle}

\documentstyle[11pt]{article}
\input amssym.def
\begin{document}
\baselineskip=18pt \setcounter{page}{1}
\begin{center}
{\LARGE \bf On PPT States in ${\cal C}^K \otimes {\cal C}^M\otimes
{\cal C}^N$ Composite\\ Quantum Systems}
\end{center}
\vskip 2mm

\begin{center}
{\normalsize Xiao-Hong Wang$^1$, Shao-Ming Fei$^{1,\  2}$,  Zhi-Xi
Wang$^1$, and Ke Wu$^1$  }
\medskip

\begin{minipage}{4.8in}
{\small \sl $ ^1$
Department of Mathematics, Capital  Normal University, Beijing,
China}\\
{\small \sl $ ^2$ Institute of Applied Mathematics, University of
Bonn,  53115 Bonn, Germany}
\end{minipage}
\end{center}

\vskip 2mm
\begin{center}

\begin{minipage}{4.8in}

\centerline{\bf Abstract}
\bigskip

We study the general representations of positive partial transpose
(PPT) states in ${\cal C}^K \otimes {\cal C}^M \otimes {\cal
C}^N$. For the PPT states with rank-$N$ a canonical form is
obtained, from which a sufficient separability condition is
presented.

\vskip 1mm
PACS number(s): 03.67.Hk, 03.65.Ta, 89.70.+c
\end{minipage}
\end{center}

\vskip0.4cm

Due to the importance of quantum entangled states in quantum
information and computation
\cite{DiVincenzo,teleport,densecrypto}, much effort has been done
recently towards an operational characterization of separable
states \cite{Werner,AB3H,BCH}. The manifestations of mixed-state
entanglement can be very subtle \cite{10}. Till now
there is no general efficient criterion in judging the
separability. The Bell inequalities
\cite{Bell64}, Peres PPT criterion \cite{Peres96a}, reduction
criterion \cite{2hPRA99,cag99}, majorization \cite{nielson01},
entanglement witnesses \cite{3hPLA223,ter00}, extension of Peres
criterion \cite{dps}, matrix realignment \cite{ru02}, generalized
partial transposition criterion (GPT) \cite{chenPLA02},
generalized reduced criterion \cite{grc}, give some necessary (and
also sufficient for some special cases \cite{3hPLA223}) conditions
for separability. The separability criterion in \cite{3hPLA223} is
both necessary and sufficient but not operational. For low rank
density matrices there are also some necessary and sufficient
operational criteria of separability \cite{hlpre,hlpre1}.

In \cite{22n,222n} the separability and entanglement of quantum mixed states
in ${\cal C}^2\otimes {\cal C}^2\otimes {\cal C}^N$,
${\cal C}^2\otimes {\cal C}^3\otimes {\cal C}^N$ and
${\cal C}^2\otimes{\cal C}^2\otimes {\cal C}^2\otimes {\cal C}^N$ composite
quantum systems have been studied in terms
of matrix analysis on tensor spaces. It is shown that
all such quantum states $\rho$ with positive partial transposes
and rank $r(\rho)\leq N$ are separable.
In this article we extend the results in \cite{22n}
to the case of composite quantum systems in ${\cal C}^K \otimes
{\cal C}^M\otimes {\cal C}^N$ with general dimensions $K,\,M,\,N\in\Nb$.
We give a canonical form of PPT states in ${\cal C}^K \otimes
{\cal C}^M\otimes {\cal C}^N$ with rank $N$ and present a
sufficient separability criterion.

A separable state in ${\cal C}^K_{\sc a} \otimes {\cal C}^M_{\sc
b}  \otimes {\cal C}^N_{\sc c}$ is of the form:
\begin{equation}
\rho _{\sc abc}=\sum_{i}p_{i}\rho^{i}_{\sc a}\otimes \rho^{i}_{\sc
b} \otimes \rho^{i}_{\sc c}, \label{sep}
\end{equation}
where $\sum_{i}p_{i}=1$, $0< p_i\leq 1$, $\rho^{i}_{\alpha}$ are
density matrices associated with the subsystems $\alpha$,
$\alpha={\sc a,b,c}$. In the following we denote by $R(\rho)$,
$K(\rho)$, $r(\rho)$ and $k(\rho)$ the range, kernel, rank and the
dimension of the kernel of $\rho$, respectively.

We first derive a canonical form of PPT states in ${\cal
C}^3_{\sc a} \otimes {\cal C}^3_{\sc b} \otimes {\cal C}^N_{\sc
c}$ with rank $N$, which allows for an explicit decomposition of a
given state in terms of convex sum of projectors on product
vectors. Let $|0_{\sc a}\ra$, $|1_{\sc a}\ra$, $|2_{\sc a}\ra$;
$|0_{\sc b}\ra$, $|1_{\sc b}\ra$, $|2_{\sc b}\ra$; and $|0_{\sc
c}\ra\, \cdots \,|N-1_{\sc c}\ra$ be some local bases of the
sub-systems ${\sc a,~b,~c}$ respectively.

{\bf Lemma 1. }\ \  Every PPT state $\rho$ in ${\cal C}^3_{\sc a}
\otimes {\cal C}^3_{\sc b}\otimes {\cal C}^N_{\sc c}$ such that
$r(\la 2_{\sc a}, 2_{\sc b}|\rho |2_{\sc a}, 2_{\sc
b}\ra)=r(\rho)=N$, can be transformed into the following canonical
form by using a reversible local operation: \be \rho=\sqrt{F}[DB\
\ DA\ \ D\ \ CB\ \ CA\ \ C\ B\ \ A\ \ I]^{\dag} [DB\ \ DA\ \ D\ \
CB\ \ CA\ \ C\ \ B\ \ A\ \ I]\sqrt{F} \ee where $A$, $B$, $C$,
$D$, $F$ and the identity $I$ are $N\times N$ matrices acting on
${\cal C}_{\sc c}^N$ and satisfy the following relations: $[A,\
A^{\dag}]=[B,\ B^{\dag}]=[C,\ C^{\dag}]=[D,\ D^{\dag}]=[B,\
A]=[B,\ A^{\dag}] =[C,\ A]=[C,\ A^{\dag}]=[D,\ A]=[D,\
A^{\dag}]=[C,\ B]=[C,\ B^{\dag}]=[D,\ B]=[D,\ B^{\dag}]= [D,\
C]=[D,\ C^{\dag}]=0$ and $F=F^{\dag}$ ($\dag$ stands for the
transposition and conjugate).

{\bf Proof.}\ In the basis we considered, a density matrix $\rho$ in
${\cal C}^3_{\sc a} \otimes {\cal C}^3_{\sc b} \otimes {\cal C}^N_{\sc c}$ with
rank $N$ can be always written as:
\be\label{r}
\rho=\left(
\begin{array}{ccccccccc}
E_1&E_{12}&E_{13}&E_{14}&E_{15}&E_{16}&E_{17}&E_{18}&E_{19}\\
E_{12}^{\dag}&E_2&E_{23}&E_{24}&E_{25}&E_{26}&E_{27}&E_{28}&E_{29}\\
E_{13}^{\dag}&E_{23}^{\dag}&E_{3}&E_{34}&E_{35}&E_{36}&E_{37}&E_{38}&E_{39}\\
E_{14}^{\dag}&E_{24}^{\dag}&E_{34}^{\dag}&E_4&E_{45}&E_{46}&E_{47}&E_{48}&E_{49}\\
E_{15}^{\dag}&E_{25}^{\dag}&E_{35}^{\dag}&
E_{45}^{\dag}&E_5&E_{56}&E_{57}&E_{58}&E_{59}\\
E_{16}^{\dag}&E_{26}^{\dag}&E_{36}^{\dag}&
E_{46}^{\dag}&E_{56}^{\dag}&E_6&E_{67}&E_{68}&E_{69}\\
E_{17}^{\dag}&E_{27}^{\dag}&E_{37}^{\dag}&
E_{47}^{\dag}&E_{57}^{\dag}&E_{67}^{\dag}&E_7&E_{78}&E_{79}\\
E_{18}^{\dag}&E_{28}^{\dag}&E_{38}^{\dag}&
E_{48}^{\dag}&E_{58}^{\dag}&E_{68}^{\dag}&E_{78}^{\dag}&E_8&E_{89}\\
E_{19}^{\dag}&E_{29}^{\dag}&E_{39}^{\dag}&
E_{49}^{\dag}&E_{59}^{\dag}&E_{69}^{\dag}&E_{79}^{\dag}&E_{89}^{\dag}&E_9\\
\end{array}
\right),
\ee
where $E's$ are $N \times N$ matrices, $r(E_9)=N$.
The projection $\la 2_{\sc a}|\rho|2_{\sc a}\ra$ gives rise to a state
\be \tilde{\rho}=
\la2_A|\rho|2_A\ra
=\left(
\begin{array}{ccc}
E_7&E_{78}&E_{79}\\
E_{78}^{\dag}&E_8&E_{89}\\
E_{79}^{\dag}&E_{89}^{\dag}&E_9
\end{array}
\right), \ee which is a state in ${\cal C}^3_{\sc b} \otimes {\cal
C}^N_{\sc c}$ with $r(\tilde{\rho})=r(\rho)=N$. Let $t_\alpha$
denote the partial transposition with respect to the subsystem
$\alpha$. As every principal minor determinant of $\tilde
\rho^{t_{\sc b}}$ ($\tilde \rho^{t_{\sc c}}$) is some principal
minor determinant of $\rho$, the fact that $\rho$ is PPT implies
that $\tilde \rho$ is also PPT, i.e., $\tilde{\rho}\ge 0$. After
performing a  reversible local non-unitary "filtering"
$\frac{1}{\sqrt{E_9}}$ on the third system and  using Lemma 4
in \cite{hlpre} the matrix $\tilde{\rho}$ can be written as
 \be \tilde{\rho}=\left(
\begin{array}{ccc}
B^{\dag}B&B^{\dag}A&B^{\dag}\\
A^{\dag}B&A^{\dag}A&A^{\dag}\\ B&A&I
\end{array}
\right),
\ee
where $[A,A^{\dag}]=[B,B^{\dag}]=[B,A]=[B,A^{\dag}]=0$.

Similarly, if we consider the projection $\la 2_{\sc
b}|\rho|2_{\sc b}\ra$ , for the same reasons as above we conclude
that the resulting matrix
$$
\ba{rcl}
\bar{\rho}&=&\la2_B|\rho|2_B\ra\\[5mm]
&=&\left(
\begin{array}{ccc}
E_3&E_{36}&E_{39}\\
E_{36}^{\dag}&E_6&E_{69}\\
E_{39}^{\dag}&E_{69}^{\dag}&E_9
\end{array}
\right) =\left(
\begin{array}{ccc}
D^{\dag}D&D^{\dag}C&D^{\dag}\\
C^{\dag}D&C^{\dag}C&C^{\dag}\\ D&C&I
\end{array}
\right), \ea
$$
where $[C,C^{\dag}]=[D,D^{\dag}]=[D,C]=[D,C^{\dag}]=0$.

Summarizing, after performing a local filtering operation
$\frac{1}{\sqrt{E_9}}$ we can bring the matrix $\rho$ to the form:
\be\label{rm} \rho=\left(
\begin{array}{ccccccccc}
E_1&E_{12}&E_{13}&E_{14}&E_{15}&E_{16}&E_{17}&E_{18}&E_{19}\\
E_{12}^{\dag}&E_2&E_{23}&E_{24}
&E_{25}&E_{26}&E_{27}&E_{28}&E_{29}\\
E_{13}^{\dag}&E_{23}^{\dag}&D^{\dag}D&E_{34}
&E_{35}&D^{\dag}C&E_{37}&E_{38}&D^{\dag}\\
E_{14}^{\dag}&E_{24}^{\dag}&E_{34}^{\dag}&E_4&E_{45}
&E_{46}&E_{47}&E_{48}&E_{49}\\
E_{15}^{\dag}&E_{25}^{\dag}&E_{35}^{\dag}&E_{45}^{\dag}& E_5
&E_{56}&E_{57}&E_{58}&E_{59}\\
E_{16}^{\dag}&E_{26}^{\dag}&C^{\dag}D&E_{46}^{\dag}&
E_{56}^{\dag}&C^{\dag}C&E_{67}&E_{68}&C^{\dag}\\
E_{17}^{\dag}&E_{27}^{\dag}&E_{37}^{\dag}&E_{47}^{\dag}&
E_{57}^{\dag}&E_{67}^{\dag}&B^{\dag}B&B^{\dag}A&B^{\dag}\\
E_{18}^{\dag}&E_{28}^{\dag}&E_{38}^{\dag}&E_{48}^{\dag}&
E_{58}^{\dag}&E_{68}^{\dag}&A^{\dag}B&A^{\dag}A&A^{\dag}\\
E_{19}^{\dag}&E_{29}^{\dag}&D&E_{49}^{\dag}&E_{59}^{\dag}&C&B&A&I\\
\end{array}
\right).
\ee
 Notice that $\la \Psi_f|\rho|\Psi_f\ra=0$ for
$|\Psi_f\ra=|21\ra |f\ra -|22\ra A|f\ra$ and arbitrary $|f\ra\in
{\cal C}^N_{\sc C}$. As $\rho\geq 0$ we have that $|\Psi_f\ra$
is in the kernel. Using the same method , we can get that $\rho$
has the following kernel vectors: \be \ba{ll} |21\ra|f\ra-|22\ra
A|f\ra,&~~
|20\ra|g\ra-|22\ra B|g\ra,\\[3mm]
|12\ra|h\ra-|22\ra C|h\ra,&~~ |02\ra|k\ra-|22\ra D|k\ra, \ea \ee
for all vectors $|f\ra,~|g\ra,~|h\ra,~|k\ra\in{\cal C}^N_{\sc c}$.
This implies
\be\label{imp} \ba{lll} E_{38}=D^{\dag}A,&~~
E_{68}=C^{\dag}A,&~~ E_{37}=D^{\dag}B,\\[3mm]
 E_{67}=C^{\dag}B,&~~
E_{13}=E_{19}D,&~~ E_{23}=E_{29}D,\\[3mm]
 E_{34}^{\dag}=E_{49}D,&~~
E_{35}^{\dag}=E_{59}D,&~~
 E_{i6}=E_{i9}C,\\[3mm]
  E_{i7}=E_{i9}B,&~~
E_{i8}=E_{i9}A,&~~i=1,2,4,5.\\[3mm]
\ea \ee

Substituting (\ref{imp}) into (\ref{rm}) and considering partial
transposition of $\rho$ with respect to the first sub-system ${\sc
a}$, we have \be \rho^{t_{\sc a}}=\left(
\begin{array}{ccccccccc}
E_1&E_{12}&E_{13}&E_{14}^{\dag}&E_{24}^{\dag}&E_{49}D&B^{\dag}E_{19}^{\dag}
&B^{\dag}E_{29}^{\dag}&B^{\dag}D\\
E_{12}^{\dag}&E_2&E_{23}&E_{15}^{\dag}&
E_{25}^{\dag}&E_{59}D&A^{\dag}E_{19}^{\dag}&A^{\dag}E_{29}^{\dag}&A^{\dag}D\\
E_{13}^{\dag}&E_{23}^{\dag}&D^{\dag}D&
C^{\dag}E_{19}^{\dag}&C^{\dag}E_{29}^{\dag}&C^{\dag}D&E_{19}^{\dag}&E_{29}^{\dag}&D\\
E_{14}&E_{15}&E_{19}C&E_4&E_{45}&E_{49}C
&B^{\dag}E_{49}^{\dag}&B^{\dag}E_{59}^{\dag}&B^{\dag}C\\
E_{24}&E_{25}&E_{29}C&E_{45}^{\dag}&E_5&E_{59}C
&A^{\dag}E_{49}^{\dag}&A^{\dag}E_{59}^{\dag}&A^{\dag}C\\
E_{34}&E_{35}&D^{\dag}C&C^{\dag}E_{49}^{\dag}&C^{\dag}E_{59}^{\dag}
&C^{\dag}C&E_{49}^{\dag}&E_{59}^{\dag}&C\\
E_{19}B&E_{19}A&E_{19}&E_{49}B&E_{49}A&E_{49}
&B^{\dag}B&B^{\dag}A&B^{\dag}\\
E_{29}B&E_{29}A&E_{29}&E_{59}B&E_{59}A&E_{59}
&A^{\dag}B&A^{\dag}A&A^{\dag}\\
D^{\dag}B&D^{\dag}A&D^{\dag}&C^{\dag}B&C^{\dag}A&C^{\dag}&B&A&I
\end{array}
\right).
\ee

Since the partial transposition with respect to the sub-system
${\sc a}$ is positive, $\rho^{t_{\sc a}}\ge 0$, and it does not
change $\la 2_{\sc a}|\rho|2_{\sc a}\ra$, we still have
$|20\ra|g\ra-|22\ra B|g\ra$, $|21\ra|f\ra-|22\ra A|f\ra\in
k(\rho^{t_{\sc a}})$. This gives rise to the following equalities:
\be\label{imp1} \ba{lll} E_{19}=B^{\dag}D^{\dag},&~~
E_{29}=A^{\dag}D^{\dag},&~~\\[3mm] E_{49}=B^{\dag}C^{\dag},&~~
E_{59}=A^{\dag}C^{\dag}.\\[3mm]
 \ea \ee
 $\rho$ is then of the following form:
\be \left(
\begin{array}{ccccccccc}
E_1\!&\!E_{12}\!&\!B^{\dag}D^{\dag}D\!&\!E_{14}\!&\!E_{15}\!&\!
B^{\dag}D^{\dag}C\!&\!B^{\dag}D^{\dag}B \!&\!B^{\dag}D^{\dag}A
\!&\!B^{\dag}D^{\dag}\\
E_{12}^{\dag}\!&\!E_2\!&\!A^{\dag}D^{\dag}D\!&\!E_{24}\!&\!E_{25}\!&\!
A^{\dag}D^{\dag}C\!&\!A^{\dag}D^{\dag}B\!&\!A^{\dag}D^{\dag}A\!&\!A^{\dag}D^{\dag}\\
D^{\dag}DB\!&\!D^{\dag}DA\!&\!D^{\dag}D\!&\!
D^{\dag}CB\!&\!D^{\dag}CA\!&\!D^{\dag}C\!&\!D^{\dag}B\!&\!D^{\dag}A\!&\!D^{\dag}\\
E_{14}^{\dag}\!&\!E_{24}^{\dag}\!&\!B^{\dag}C^{\dag}D\!&\!E_4\!&\!E_{45}\!&\!
B^{\dag}C^{\dag}C\!&\!B^{\dag}C^{\dag}B\!&\!B^{\dag}C^{\dag}A\!&\!B^{\dag}C^{\dag}\\
E_{15}^{\dag}\!&\!E_{25}^{\dag}\!&\!A^{\dag}C^{\dag}D\!&\!E_{45}^{\dag}\!&\!
E_5\!&\!A^{\dag}C^{\dag}C\!&\!A^{\dag}C^{\dag}B\!&\!A^{\dag}C^{\dag}A
\!&\!A^{\dag}C^{\dag}\\
C^{\dag}DB\!&\!C^{\dag}DA\!&\!C^{\dag}D\!&\!
C^{\dag}CB\!&\!C^{\dag}CA\!&\!C^{\dag}C\!&\!C^{\dag}B\!&\!C^{\dag}A\!&\!C^{\dag}\\
B^{\dag}DB\!&\!B^{\dag}DA\!&\!B^{\dag}D\!&\!
B^{\dag}CB\!&\!B^{\dag}CA\!&\!B^{\dag}C\!&\!B^{\dag}B\!&\!B^{\dag}A\!&\!B^{\dag}\\
A^{\dag}DB\!&\!A^{\dag}DA\!&\!A^{\dag}D\!&\!
A^{\dag}CB\!&\!A^{\dag}CA\!&\!A^{\dag}C\!&\!A^{\dag}B\!&\!A^{\dag}A\!&\!A^{\dag}\\
DB\!&\!DA\!&\!D\!&\!CB\!&\!CA\!&\!C\!&\!B\!&\!A\!&\!I
\end{array}
\right).\\
\ee
Set
$$
X=\left(
\begin{array}{ccccc}
E_{15}&B^{\dag}D^{\dag}C&B^{\dag}D^{\dag}B&B{\dag}D^{\dag}A&B^{\dag}D^{\dag}\\
E_{25}&A^{\dag}D^{\dag}C&A^{\dag}D^{\dag}B&A^{\dag}D^{\dag}A&A^{\dag}D^{\dag}\\
D^{\dag}CA&D^{\dag}C&D^{\dag}B&D^{\dag}A&D^{\dag}\\
E_{45}&B^{\dag}C^{\dag}C&B^{\dag}C^{\dag}B&B^{\dag}C^{\dag}A&B^{\dag}C^{\dag}\\
\end{array}
\right),
$$

$$
Y=\left(
\begin{array}{cccc}
E_1&E_{12}&B^{\dag}D^{\dag}D&E_{14}\\
E_{12}^{\dag}&E_2&A^{\dag}D^{\dag}D&E_{24}\\
D^{\dag}DB&D^{\dag}DA&D^{\dag}D&D^{\dag}CB\\
E_{14}^{\dag}&E_{24}^{\dag}&B^{\dag}C^{\dag}D&E_4\\
\end{array}
\right)
$$
and
$$
\rho_5=\Sigma+{\rm diag}(\Delta,\ 0,\ 0,\ 0,\ 0),
$$ \\
where
$$
\Sigma=\left(
\begin{array}{ccccc}
A^{\dag}C^{\dag}CA&A^{\dag}C^{\dag}C&
A^{\dag}C^{\dag}B&A^{\dag}C^{\dag}A&A^{\dag}C^{\dag}\\
C^{\dag}CA&C^{\dag}C&C^{\dag}B&C^{\dag}A&C^{\dag}\\
B^{\dag}CA&B^{\dag}C&B^{\dag}B&B^{\dag}A&B^{\dag}\\
A^{\dag}CA&A^{\dag}C&A^{\dag}B&A^{\dag}A&A^{\dag}\\
CA&C&B&A&I
\end{array}
\right),
$$
\be \label{delta} \Delta=E_5-A^{\dag}C^{\dag}CA \ee
and diag$(A_1,A_2,...,A_m)$ denotes a diagonal block matrix with
blocks $A_1,A_2,...,A_m$. $\rho$ can then be written in the
following partitioned matrix form:
\[
\rho=\left(
\begin{array}{cc}
Y&X\\
X^{\dag}&\rho_5
\end{array}
\right).
\]

As $\Sigma$ possesses the following  4N kernel vectors:
$$
\begin{array}{ll}
&(\la f|, \ 0,\ 0,\ 0,-\la f|A^\dag C^\dag )^T,\ \
(0, \la g|,0,\ 0, -\la g|C^\dag )^T,\\[3mm]
&(0,\ 0,\la h|, \ 0,\ -\la h|B^\dag )^T,\ \ (0,\ 0,\ 0,\la i|,\
-\la i|A^\dag )^T
\end{array}
$$
for arbitrary $|f\ra,\ |g\ra,\ |h\ra,\ |i\ra\in{\cal C}^N_{\sc c}$,
the kernel $K(\Sigma)$ has at least dimension $4N$. On the
other hand $r(\Sigma)+k(\Sigma)=5N$, therefore $r(\Sigma)\leq N$.
As the range of $\Sigma$ has at least dimension $N$ due to the
identity entry on the diagonal, we have $r(\Sigma)=N$. Notice
that $r(\rho_5)\leq r(\rho)=N$, it is easy to see that
$r(\rho_5)=N$. To show that $\Delta= 0$, we make the following
elementary row transformations on the matrix $\rho_5$,
\be\label{add} \left(
\begin{array}{ccccc}
I&0&0&0&-A^{\dag}C^{\dag}\\
0&I&0&0&-C^{\dag}\\
0&0&I&0&-B^{\dag}\\
0&0&0&I&-A^{\dag}\\
0&0&0&0&I
\end{array}
\right)\rho_5= \left(
\begin{array}{ccccc}
\Delta&0&0&0&0\\
0&0&0&0&0\\
0&0&0&0&0\\
0&0&0&0&0\\
CA&C&B&A&I
\end{array}
\right). \ee As the rank of $\rho_5$ is $N$, from (\ref{add}) we
have $\Delta=0$, and hence  $E_5=A^{\dag}C^{\dag}CA.$

Now, notice that $\la \Psi_f|\rho|\Psi_f\ra=0$ for
$|\Psi_f\ra=|11\ra |f\ra -|22\ra CA|f\ra$ and arbitrary
$|f\ra\in{\cal C}^N_{\sc c}$. Since $\rho\geq 0$ we have $0=\rho
|\Psi_f\ra=|00\ra (E_{15}-B^{\dag}D^{\dag}CA) |f\ra +|01\ra
(E_{25}-A^{\dag}D^{\dag}CA )|f\ra +|10\ra
(E_{45}-B^{\dag}C^{\dag}CA)|f\ra$, which, as $|f\ra$ is arbitray,
leads to $E_{15}=B^{\dag}D^{\dag}CA$, $E_{25}=A^{\dag}D^{\dag}CA$,
$E_{45}=B^{\dag}C^{\dag}CA$, thus the matrix $\rho$ becomes \be
\left(
\begin{array}{ccccccccc}
E_1\!&\!E_{12}\!&\! B^{\dag}D^{\dag}D\!&\!
E_{14}\!&\!B^{\dag}D^{\dag}CA
\!&\!B^{\dag}D^{\dag}C\!&\!B^{\dag}D^{\dag}B
\!&\!B^{\dag}D^{\dag}A\!&\!B^{\dag}D^{\dag}\\
E_{12}^{\dag}&E_2&A^{\dag}D^{\dag}D&E_{24}\!&\!
A^{\dag}D^{\dag}CA\!&\!A^{\dag}D^{\dag}C\!&\!A^{\dag}D^{\dag}B
\!&\!A^{\dag}D^{\dag}A\!&\!A^{\dag}D^{\dag}\\
D^{\dag}DB\!&\!D^{\dag}DA\!&\!D^{\dag}D\!&\!
D^{\dag}CB\!&\!D^{\dag}CA\!&\!D^{\dag}C\!&\!D^{\dag}B\!&\!D^{\dag}A
\!&\!D^{\dag}\\
E_{14}^{\dag}\!&\!E_{24}^{\dag}\!&\!B^{\dag}C^{\dag}D\!&\!E_4\!&\!
B^{\dag}C^{\dag}CA\!&\!B^{\dag}C^{\dag}C\!&\!B^{\dag}C^{\dag}B\!&\!
B^{\dag}C^{\dag}A\!&\!B^{\dag}C^{\dag}\\
A^{\dag}C^{\dag}DB\!&\!A^{\dag}C^{\dag}DA\!&\!A^{\dag}C^{\dag}D\!&\!A^{\dag}C^{\dag}CB\!&\!
A^{\dag}C^{\dag}CA\!&\!A^{\dag}C^{\dag}C\!&\!A^{\dag}C^{\dag}B\!&\!A^{\dag}C^{\dag}A
\!&\!A^{\dag}C^{\dag}\\
C^{\dag}DB\!&\!C^{\dag}DA\!&\!C^{\dag}D\!&\!C^{\dag}CB\!&\!
C^{\dag}CA\!&\!C^{\dag}C\!&\!C^{\dag}B\!&\!C^{\dag}A
\!&\!C^{\dag}\\
B^{\dag}DB\!&\!B^{\dag}DA\!&\!B^{\dag}D\!&\!B^{\dag}CB\!&\!
B^{\dag}CA\!&\!B^{\dag}C\!&\!B^{\dag}B\!&\!B^{\dag}A\!&\!B^{\dag}\\
A^{\dag}DB\!&\!A^{\dag}DA\!&\!A^{\dag}D\!&\!A^{\dag}CB\!&\!
A^{\dag}CA\!&\!A^{\dag}C\!&\!A^{\dag}B\!&\!A^{\dag}A\!&\!A^{\dag}\\
DB\!&\!DA\!&\!D\!&\!CB\!&\!CA\!&\!C\!&\!B\!&\!A\!&\!I
\end{array}
\right).\\
\ee

Similarly, we can derive $E_4=B^{\dag}C^{\dag}CB$,
$E_{14}=B^{\dag}D^{\dag}CB$, $E_{24}=A^{\dag}D^{\dag}CB$,
$E_2=A^{\dag}D^{\dag}DA$, $E_{12}=B^{\dag}D^{\dag}DA$,
$E_1=B^{\dag}D^{\dag}DB$. $\rho$ then is of the following form:
$$
\ba{l} \left(
\begin{array}{ccccccccc}
B^{\dag}D^{\dag}DB&B^{\dag}D^{\dag}DA&B^{\dag}D^{\dag}D&B^{\dag}D^{\dag}CB&
B^{\dag}D^{\dag}CA\!&\!B^{\dag}D^{\dag}C\!&\!B^{\dag}D^{\dag}B\!&\!B^{\dag}D^{\dag}A
\!&\!B^{\dag}D^{\dag}\\
A^{\dag}D^{\dag}DB&A^{\dag}D^{\dag}DA\!&\!A^{\dag}D^{\dag}D\!&\!A^{\dag}D^{\dag}CB\!&\!
A^{\dag}D^{\dag}CA\!&\!A^{\dag}D^{\dag}C\!&\!A^{\dag}D^{\dag}B\!&\!A^{\dag}D^{\dag}A
\!&\!A^{\dag}D^{\dag}\\
D^{\dag}DB&D^{\dag}DA\!&\!D^{\dag}D\!&\!D^{\dag}CB\!&\!
D^{\dag}CA\!&\!D^{\dag}C\!&\!D^{\dag}B\!&\!D^{\dag}A\!&\!D^{\dag}\\
B^{\dag}C^{\dag}DB\!&\!B^{\dag}C^{\dag}DA\!&\!B^{\dag}C^{\dag}D\!&\!
B^{\dag}C^{\dag}CB\!&\!B^{\dag}C^{\dag}CA
\!&\!B^{\dag}C^{\dag}C&B^{\dag}C^{\dag}B\!&\!B^{\dag}C^{\dag}A
\!&\!B^{\dag}C^{\dag}\\
A^{\dag}C^{\dag}DB\!&\!A^{\dag}C^{\dag}DA\!&\!A^{\dag}C^{\dag}D\!&\!
A^{\dag}C^{\dag}CB\!&\!A^{\dag}C^{\dag}CA
\!&\!A^{\dag}C^{\dag}C&A^{\dag}C^{\dag}B\!&\!A^{\dag}C^{\dag}A
\!&\!A^{\dag}C^{\dag}\\
C^{\dag}DB\!&\!C^{\dag}DA\!&\!C^{\dag}D\!&\!C^{\dag}CB\!&\!
C^{\dag}CA\!&\!C^{\dag}C\!&\!C^{\dag}B\!&\!C^{\dag}A
\!&\!C^{\dag}\\
B^{\dag}DB\!&\!B^{\dag}DA\!&\!B^{\dag}D\!&\!B^{\dag}CB\!&\!
B^{\dag}CA\!&\!B^{\dag}C\!&\!B^{\dag}B\!&\!B^{\dag}A\!&\!B^{\dag}\\
A^{\dag}DB\!&\!A^{\dag}DA\!&\!A^{\dag}D\!&\!A^{\dag}CB\!&\!
A^{\dag}CA\!&\!A^{\dag}C\!&\!A^{\dag}B\!&\!A^{\dag}A\!&\!A^{\dag}\\
DB\!&\!DA\!&\!D\!&\!CB\!&\!CA\!&\!C\!&\!B\!&\!A\!&\!I
\end{array}
\right)\\[20mm]
=[\begin{array}{ccccccccc}
DB&DA&D&CB&CA&C&B&A&I\end{array}]^{\dag}
[DB\,\,\, DA\,\,\, D\,\,\, CB\,\,\, CA\,\,\, C\,\,\, B\,\,\, A\,\,\, I]\\[3mm]
\ea
$$

The commutative relations $[A,\ D]=[B,\ D]=[A,\ C]=[B,\ C]=[A,\
D^{\dag}]=[B,\ D^{\dag}]=[A,\ C^{\dag}]=[B,\ C^{\dag}]=0$ follow
from the positivity of all partial transpositions of $\rho$. We
first consider: $$ \rho^{t_{\sc b}}=\left(
\begin{array}{ccccccccc}
B^{\dag}D^{\dag}DB\!&\!A^{\dag}D^{\dag}DB\!&\!D^{\dag}DB\!&\!
B^{\dag}D^{\dag}CB\!&\!A^{\dag}D^{\dag}CB
\!&\!D^{\dag}CB\!&\!B^{\dag}D^{\dag}B
\!&\!A^{\dag}D^{\dag}B\!&\!D^{\dag}B\\
B^{\dag}D^{\dag}DA\!&\!A^{\dag}D^{\dag}DA\!&\!D^{\dag}DA\!&\!
B^{\dag}D^{\dag}CA\!&\!A^{\dag}D^{\dag}CA\!&\!D^{\dag}CA\!&\!B^{\dag}D^{\dag}A
\!&\!A^{\dag}D^{\dag}A\!&\!D^{\dag}A\\
B^{\dag}D^{\dag}D\!&\!A^{\dag}D^{\dag}D\!&\!D^{\dag}D\!&\!B^{\dag}D^{\dag}C\!&\!
A^{\dag}D^{\dag}C\!&\!D^{\dag}C\!&\!B^{\dag}D^{\dag}\!&\!A^{\dag}D^{\dag}
\!&\!D^{\dag}\\
B^{\dag}C^{\dag}DB\!&\!A^{\dag}C^{\dag}DB\!&\!C^{\dag}DB\!&\!
B^{\dag}C^{\dag}CB\!&\!A^{\dag}C^{\dag}CB\!&\!C^{\dag}CB\!&\!
B^{\dag}C^{\dag}B\!&\!A^{\dag}C^{\dag}B\!&\!C^{\dag}B\\
B^{\dag}C^{\dag}DA\!&\!A^{\dag}C^{\dag}DA\!&\!C^{\dag}DA\!&\!
B^{\dag}C^{\dag}CA\!&\!A^{\dag}C^{\dag}CA\!&\!C^{\dag}CA\!&\!
B^{\dag}C^{\dag}A\!&\!A^{\dag}C^{\dag}A\!&\!C^{\dag}A\\
B^{\dag}C^{\dag}D\!&\!A^{\dag}C^{\dag}D\!&\!C^{\dag}D\!&\!
B^{\dag}C^{\dag}C\!&\!A^{\dag}C^{\dag}C\!&\!C^{\dag}C\!&\!B^{\dag}C^{\dag}
\!&\!A^{\dag}C^{\dag}\!&\!C^{\dag}\\
B^{\dag}DB\!&\!A^{\dag}DB\!&\!DB\!&\!B^{\dag}CB\!&\!
A^{\dag}CB\!&\!CB\!&\!B^{\dag}B\!&\!A^{\dag}B\!&\!B\\
B^{\dag}DA\!&\!A^{\dag}DA\!&\!DA\!&\!B^{\dag}CA\!&\!
A^{\dag}CA\!&\!CA\!&\!B^{\dag}A\!&\!A^{\dag}A\!&\!A\\
B^{\dag}D\!&\!A^{\dag}D\!&\!D\!&\!B^{\dag}C\!&\!A^{\dag}C\!&\!C\!&\!B^{\dag}\!&\!A^{\dag}\!&\!I
\end{array}
\right)
$$
Due to the positivity, the matrix $\rho^{t_B}$ must
possess the kernel vector $|12\ra |f\ra -|22\ra C |f\ra$, $|02\ra
|g\ra -|22\ra D |g\ra$, which implies that $[A,\ C]=[B,\ C]=[A,\
D]=[B,\ D]=0$. The matrix $\rho^{t_B}$ can be then written as:
\[
\rho^{t_B}=\left(
\begin{array}{c}
D^{\dag}B\\
D^{\dag}A\\
D^{\dag}\\
C^{\dag}B\\
C^{\dag}A\\
C^{\dag}\\
C\\
B\\
A\\
I
\end{array}
\right) \left(
\begin{array}{ccccccccc}
B^{\dag}D&A^{\dag}D&D&B^{\dag}C&A^{\dag}C&C&B^{\dag}&A^{\dag}&I
\end{array}
\right),
\]
which implies automatically the positivity.

From the positivity of $\rho^{t_{AB}}$,
 \[
\rho^{t_{AB}}=\left(
\begin{array}{ccccccccc}
B^{\dag}D^{\dag}DB&A^{\dag}D^{\dag}DB&D^{\dag}DB&B^{\dag}C^{\dag}DB
&A^{\dag}C^{\dag}DB&C^{\dag}DB&B^{\dag}DB&A^{\dag}DB&DB\\
B^{\dag}D^{\dag}DA&A^{\dag}D^{\dag}DA&D^{\dag}DA&B^{\dag}C^{\dag}DA
&A^{\dag}C^{\dag}DA&C^{\dag}DA&B^{\dag}DA&A^{\dag}DA&DA\\
B^{\dag}D^{\dag}D&A^{\dag}D^{\dag}D&D^{\dag}D&B^{\dag}C^{\dag}D
&A^{\dag}C^{\dag}D&C^{\dag}D&B^{\dag}D&A^{\dag}D&D\\
B^{\dag}D^{\dag}CB&A^{\dag}D^{\dag}CB&D^{\dag}CB&B^{\dag}C^{\dag}CB
&A^{\dag}C^{\dag}CB&C^{\dag}CB&B^{\dag}CB&A^{\dag}CB&CB\\
B^{\dag}D^{\dag}CA&A^{\dag}D^{\dag}CA&D^{\dag}CA&B^{\dag}C^{\dag}CA
&A^{\dag}C^{\dag}CA&C^{\dag}CA&B^{\dag}CA&A^{\dag}CA&CA\\
B^{\dag}D^{\dag}C&A^{\dag}D^{\dag}C&D^{\dag}C&B^{\dag}C^{\dag}C
&A^{\dag}C^{\dag}C&C^{\dag}C&B^{\dag}C&A^{\dag}C&C\\
B^{\dag}D^{\dag}B&A^{\dag}D^{\dag}B&D^{\dag}B&B^{\dag}C^{\dag}B
&A^{\dag}C^{\dag}B&C^{\dag}B&B^{\dag}B&A^{\dag}B&B\\
B^{\dag}D^{\dag}A&A^{\dag}D^{\dag}A&D^{\dag}A&B^{\dag}C^{\dag}A
&A^{\dag}C^{\dag}A&C^{\dag}A&B^{\dag}A&A^{\dag}A&A\\
B^{\dag}D^{\dag}&A^{\dag}D^{\dag}&D^{\dag}&B^{\dag}C^{\dag}
&A^{\dag}C^{\dag}&C^{\dag}&B^{\dag}&A^{\dag}&I
\end{array}
\right),
\]
we have that $|12\ra |f\ra -|22\ra C^{\dag} |f\ra$, $|02\ra |g\ra
-|22\ra D^{\dag} |g\ra$ are kernel vectors, which
results in $[A,\ D^{\dag}]=[B,\ D^{\dag}]=[A,\ C^{\dag}]=[B,\
C^{\dag}]=0$. $\rho^{t_{AB}}$ is then of the form:
\[
\rho^{t_{AB}}=\left(
\begin{array}{c}
DB\\
DA\\
D\\
CB\\
CA\\
C\\
B\\
A\\
I
\end{array}
\right) \left(
\begin{array}{ccccccccc}
B^{\dag}D^{\dag}&A^{\dag}D^{\dag}&D^{\dag}&B^{\dag}C^{\dag}&A^{\dag}C^{\dag}&C^{\dag}&B^{\dag}&A^{\dag}&I
\end{array}
\right).
\]
This form assures positive definiteness, and concludes the proof
of the Lemma. $\Box$

Using Lemma 1 we can prove the following Theorem:

{\bf Theorem 1.}\ \ A PPT-state $\rho$ in ${\cal C}^3 \otimes
{\cal C}^3 \otimes {\cal C}^N$ with $r(\rho)=N$ is separable if
there exists a product basis $|e_A,\ f_B\ra$ such that $r(\la
e_A,\ f_B|\rho |e_A,\ f_B\ra)=N$.

{\bf Proof. }\ \ According to the Lemma the PPT state $\rho$ can be written as
\[
\rho=
\left(
\begin{array}{c}
B^{\dag}D^{\dag}\\A^{\dag}D^{\dag}\\
D^{\dag}\\B^{\dag}C^{\dag}\\A^{\dag}C^{\dag}\\C^{\dag}\\B^{\dag}\\A^{\dag}\\I
\end{array}
\right)
\left(
\begin{array}{ccccccccc}
DB&DA&D&CB&CA&C&B&A&I
\end{array}
\right).
\]
Since all $A$, $A^\dag$, $B$, $B^\dag$, $C$ , $C^\dag$, $D$ and
$D^\dag$ commute, they have common eigenvectors $|f_n\ra$. Let
$a_n$, $b_n$, $c_n$ and $d_n$ be the corresponding eigenvalues of
$A$, $B$, $C$ and $D$ respectively. We have
\[
\la f_n|\rho |f_n\ra=
\left(
\begin{array}{c}
b_n^*d_n^*\\[2mm]a_n^*d_n^*\\[2mm]d_n^*\\[2mm]b_n^*c_n^*\\[2mm]
a_n^*c_n^*\\[2mm]c_n^*\\[2mm]
b_n^*\\[2mm]a_n^*\\[2mm]1\\[2mm]
\end{array}
\right)
\left(
\begin{array}{ccccccccc}
d_nb_n&d_na_n&d_n&c_nb_n&c_na_n&c_n&b_n&a_n&1
\end{array}
\right)\hskip5cm
\]
\[
=\left[\left(\begin{array}{c}d_n^*\\c_n^*\\1\end{array}\right)\otimes
\left(\begin{array}{c}b_n^*\\a_n^*\\1\end{array}\right)\right]
(d_n\,\,c_n\,\, 1)\otimes(b_n\,\,a_n\,\,1)=|e_{\sc a},f_{\sc
b}\ra\la e_{\sc a},f_{\sc b}|.
\]
We can thus write $\rho$ as
$$\rho=\sum_{n=1}^N|\psi_n\ra\la \psi_n|\otimes |\phi_n\ra\la \phi_n|
\otimes |f_n\ra\la f_n|,
$$
where
$$
|\psi_n\ra=\left(\begin{array}{c}d_n^*\\
c_n^*\\1\end{array}\right),~~~
|\phi_n\ra=\left(\begin{array}{c}b_n^*\\a_n^*\\
 1\end{array}\right).
$$
Because the local transformations are reversible, we can now
apply the inverse transformations and obtain a decomposition of the initial
state $\rho$ in a sum of projectors onto product vectors. This
proves the separability of $\rho$. \hfill $\Box$

The above approach can be extended to the case of higher
dimensions like ${\cal C}^3_{\sc a} \otimes {\cal C}^M_{\sc
b}\otimes {\cal C}^N_{\sc c}$. Let $|0_{\sc a}\ra$, $|1_{\sc
a}\ra$, $|2_{\sc a}\ra$; $|0_{\sc b}\ra$, $\cdots$, $|M-1_{\sc
b}\ra$; and $|0_{\sc c}\ra\, \cdots \,|N-1_{\sc c}\ra$ be some
local bases of the sub-systems ${\sc a,~b,~c}$ respectively. From
Lemma 1 it is straightforward to prove the following conclusion:

{\bf Lemma 2. }\ \  Every PPT state $\rho$ in
${\cal C}^3_{\sc a} \otimes {\cal C}^M_{\sc b}\otimes {\cal
C}^N_{\sc c}$ such that $r(\la 2_A, M-1_B|\rho |2_A,
M-1_B\ra)=r(\rho)=N$, can be transformed into the following
canonical form by using a reversible local operation:
\be
\rho=\sqrt{F} T^{\dag}T \sqrt{F},
\ee
where $T=(C\ \ B\ \
I)\otimes (A_{M-1}\ \ A_{M-2}\
  \cdots\ \ A_1\ \ I)$, $A_i$, $B$, $C$, $F$
and the identity $I$ are $N\times N$ matrices acting on ${\cal
C}_{\sc c}^N$ and satisfy the following relations: $[A_i,\
A_j]=[A_i,\ {A_j}^{\dag}]=[B,\ B^{\dag}]=[C,\ C^{\dag}]=[B,\
A_i]=[B,\ {A_i}^{\dag}]=[C,\ A_i]=[C,\ {A_i}^{\dag}]=0$,
$i,j=1,2,\cdots,M-1$ and $F=F^{\dag}$.

Extending Theorem 1 to higher dimensions, we have:

 {\bf Theorem 2.}\ \ A PPT-state $\rho$ in ${\cal C}^3 \otimes {\cal C}^M
\otimes {\cal C}^N$ with $r(\rho)=N$ is separable if there exists
a product basis $|e_A,\ f_B\ra$ such that $r(\la e_A,\ f_B|\rho
|e_A,\ f_B\ra)=N$.

By extending Lemma 2, Theorem 2 and the results in \cite{22n} and
\cite{222n}, we can give the canonical form of PPT states in
${\cal C}^K_{\sc a} \otimes {\cal C}^M_{\sc b}\otimes {\cal
C}^N_{\sc c}$ with rank $N$. Let $|0_{\sc a}\ra$, $\cdots$,
$|K-1_{\sc a}\ra$; $|0_{\sc b}\ra$, $\cdots$, $|M-1_{\sc b}\ra$;
and $|0_{\sc c}\ra\, \cdots \,|N-1_{\sc c}\ra$ be some local bases
of the sub-systems ${\sc a,~b,~c}$ respectively.

{\bf Lemma 3. }\ \  Every PPT state $\rho$ in
${\cal C}^K_{\sc a} \otimes {\cal C}^M_{\sc b}\otimes {\cal
C}^N_{\sc c}$ such that $r(\la K-1_A, M-1_B|\rho |K-1_A,
M-1_B\ra)=r(\rho)=N$, can be transformed into the following
canonical form by using a reversible local operation:
\be\label{lemma3}\rho=\sqrt{F} T^{\dag}T \sqrt{F},\ee
where $T=(B_{K-1}\ \
B_{K-2}\ \cdots\ \ B_1\ \ I)\otimes (A_{M-1}\ \ A_{M-2}\
  \cdots\ \ A_1\ \ I)$, $A_i$,  $B_j$,
 $F$ and $I$ are $N\times N$ matrices
acting on ${\cal C}_{\sc c}^N$ and satisfy the following
relations: $[A_i,\ A_s]=[A_i,\ {A_s}^{\dag}]=[B_t,\ B_j]=[B_t,\
{B_j}^{\dag}]=[A_i,\ B_j]=[A_i,\ {B_j}^{\dag}]=0$ and
$F=F^{\dag}$, $i,s=1,2,\cdots , M-1$,\ $j,t=1,2,\cdots , K-1$.

{\bf Proof.}\ In the basis we considered, a density matrix $\rho$
in ${\cal C}^K_{\sc a} \otimes {\cal C}^M_{\sc b} \otimes {\cal
C}^N_{\sc c}$ with rank $N$ can be always written as a $KM \times
KM$ partitioned matrix. Let $E_{ij}$ be the $i,j$-element of $\rho$.
Denote $E_{ii}=E_i$. Every $E's$ are $N \times N$-matrices and
$r(E_{KM})=N$. Because $\rho$ is self-adjoint, we have
$E_{ij}=E_{ji}^{\dag},\ \ i>j$.

The projection $\la K-1_{\sc a}|\rho|K-1_{\sc a}\ra$ gives rise to
a state $ \tilde{\rho}= \la K-1_A|\rho|K-1_A\ra $ which is a state
in ${\cal C}^M_{\sc b} \otimes {\cal C}^N_{\sc c}$ with
$r(\tilde{\rho})=r(\rho)=N$. The fact that $\rho$ is PPT implies
that $\tilde \rho$ is also PPT, i.e., $\tilde{\rho}\ge 0$. Using
the Lemma 5 in \cite{hlpre} we have
\be \tilde{\rho}=
[C_1,\cdots,C_{M-1},I]^{\dag}[C_1,\cdots,C_{M-1},I],
\ee
where $[C_i,C_j^{\dag}]=[C_i,C_j]=0,\ \ i,j=1,\cdots ,M-1$.

Similarly, if we consider the projection $\la M-1_{\sc
b}|\rho|M-1_{\sc b}\ra$ , we have
\be \bar{\rho}=
[D_1,\cdots,D_{K-1},I]^{\dag}[D_1,\cdots,D_{K-1},I],
\ee
where
$[D_i,D_j^{\dag}]=[D_i,D_j]=0,\ \ i,j=1,\cdots ,K-1$. Altogether
we have $K^2+M^2-1$ $E's$.

$\rho$ has the following $M-1$ kernel vectors:
\be
|K-1,i\ra|f_i\ra-|K-1,M-1\ra C_{M-i-1}|f_i\ra,~~~~~ i=0,1,...,M-2
\ee for all vectors $|f_i\ra\in{\cal C}^N_{\sc c}$. Similarly
there are $K-1$ other kernel vectors,
\be
|j,K-1\ra|g_j\ra-|K-1,M-1\ra D_j|g_j\ra,~~~~~ j=0,1,...,K-2
\ee
for all vectors $|g_j\ra\in{\cal C}^N_{\sc c}$. From these
kernel vectors of $\rho$, we observe
that the $E_{ij}$ are dependent on the last column elements
of $\rho$. From $\rho^{t_{\sc a}}\ge 0$ and that the
partial transposition of $\rho$ with respect to the
first sub-system ${\sc a}$
does not change the positivity of $\la K-1_{\sc
a}|\rho|K-1_{\sc a}\ra$, we still have some kernel vectors that belong
to $ k(\rho^{t_{\sc a}})$, from which we can get the last column elements
and hence the last row elements of $\rho$.
Then we can write $\rho$  in the following partitioned matrix form:
\[
\rho=\left(
\begin{array}{cc}
Y&X\\
X^{\dag}&\rho_0
\end{array}
\right)
\]
with
\[
\rho_{0}=\left(
\begin{array}{cc}
E_k&Z\\
Z^{\dag}&W
\end{array}
\right),
\]
where $Z$ and $W$ are known, $k=KM-(M+1)$. Similar to the proof of
Lemma 1, denoting $\rho_0=\Sigma+{\rm diag}(\Delta,\ 0,\
0,\ldots,\ 0)$ and proving that $\Delta=0$ we get the form of $E_k$.

By repeating the procedure above, we can calculate all the
diagonal elements of $\rho$. The rest commuting relations among
$A_i$, $B_j$ can be obtained from the PPT properties of $\rho$,
similar to the case in Lemma 1. \hfill $\Box$

From the canonical form (\ref{lemma3}), we can obtain the following
result:

{\bf Theorem 3.}\ \ A PPT-state $\rho$ in ${\cal C}^K \otimes
{\cal C}^M \otimes {\cal C}^N$ with $r(\rho)=N$ is separable if
there exists a product basis $|e_A,\ f_B\ra$ such that $r(\la
e_A,\ f_B|\rho |e_A,\ f_B\ra)=N$.

In the following we give some detailed examples related to
our canonical form of PPT states and the
separability criterion.

i) An obvious separable mixed state on $K\times M\times N$ is
$ \rho=\left(
\begin{array}{cc}
{\frac{1}{N}}I&0\\
0&0
\end{array}
\right)$, where $I$ is an $N\times N$ unit matrix.
Obviously $\rho$ is a $PPT$ state with rank N, and there exist
 $|e_A\ra=|0_A\ra$, $|f_B\ra =
|0_B\ra$,  such that $\la e_A,f_B|\rho|e_A,f_B \ra ={\frac
{1}{N}}I$.  Therefore $r(\la e_A,f_B|\rho|e_A,f_B \ra) =r(\rho)
=N$. Thus the conditions in Theorem 3 are satisfied and $\rho$ is
separable. In  fact, if we set $|\psi_1\ra=|\psi_2\ra=|0_A\ra$,
$|\phi_1\ra=|\phi_2\ra=|0_B\ra$, $|\varphi_1\ra={\frac
{1}{\sqrt{2}}}(|0_C\ra+|1_C\ra)$, $|\varphi_2\ra={\frac
{1}{\sqrt{2}}}(|0_C\ra-|1_C\ra)$, $p_1=p_2={\frac {1}{2}}$, then
$\rho$ can be written in a separated form:
$$
\rho=p_1\rho_{11} \otimes \rho_{12} \otimes \rho_{13}+
     p_2\rho_{21} \otimes \rho_{22} \otimes \rho_{23},$$
where $\rho_{i1}=|\psi_i\ra\la\psi_i |,~
       \rho_{i2}=|\phi_i\ra\la\phi_i |,~
       \rho_{i3}=|\varphi_i\ra\la\varphi_i|,~
       i=1,2$.

ii) Consider a three-qubit state:
 $ \rho=\left(
\begin{array}{cc}
A&0\\
0&0
\end{array}
 \right)
$ with
 $A=\left(
\begin{array}{cc}
\frac {1}{2}&a\\
a&\frac {1}{2}
\end{array}
 \right)
$, $a\in\Rb$.$\rho$ is a mixed state as $tr{\rho}^2<1$. It is easily verified
that $\rho$ is $PPT$:
$$ \rho^{t_{A}}= \rho^{t_{B}}= \rho^{t_C}=
\rho^{t_{AB}}= \rho^{t_{AC}}= \rho^{t_{BC}}= \rho$$
and $r(\rho)=2$. Let
$|e_A\ra=|f_B\ra =|0\ra$. We have $\la
e_A,f_B|\rho|e_A,f_B \ra ={\frac {1}{2}}I$. Therefore $r(\la
e_A,f_B|\rho|e_A,f_B \ra) =r(\rho) =2$. From Theorem 3
$\rho$ is separable. In fact $\rho$ has separable form
$$
\rho=p_1\rho_{11} \otimes \rho_{12} \otimes \rho_{13}+
     p_2\rho_{21} \otimes \rho_{22} \otimes \rho_{23},$$
where $|\psi_1\ra=|\psi_2\ra =|\phi_1\ra=|\phi_2\ra=|0\ra$,
$|\varphi_1\ra={\frac {1}{\sqrt{2}}}(|0\ra+|1\ra)$,
$|\varphi_2\ra={\frac {1}{\sqrt{2}}}(|0\ra-|1\ra)$,
$p_1=p_2={\frac {1}{2}}$, and
$\rho_{i1}=|\psi_i\ra\la\psi_i |,~
       \rho_{i2}=|\phi_i\ra\la\phi_i |,~
       \rho_{i3}=|\varphi_i\ra\la\varphi_i |,~
       i=1,2$.

iii) The biseparable  three-qubit bound entangled state:
$$
\rho=\frac{1}{8}(I-\sum_{i=1}^4 |\psi_i\ra\la\psi_i|),
$$
where $|\psi_i\ra$'s are given by
$|0,1,+\ra,~|1,+,0\ra,~|+,1,0\ra,~|-,-,-\ra$
with $|\pm\ra=\frac{1}{\sqrt{2}}(|0\ra\pm|1\ra).$ $\rho$ is a
$PPT$ state as $\rho^{t_{A}}= \rho^{t_{B}}= \rho^{t_C}=
\rho^{t_{AB}}= \rho^{t_{AC}}= \rho^{t_{BC}}= \rho.$ It is
separable under any bipartite cut $A|BC, B|AC,B|CA.$ But it is
entangled (not fully separable). As $r(\rho)\not=2$ this state does
not satisfy the conditions of Theorem 3 and
the corresponding conclusions could not be deduced.

We have derived a canonical form of PPT states in ${\cal C}^K
\otimes {\cal C}^M \otimes {\cal C}^N$ with rank $N$ and a sufficient
separability criterion from this canonical form. For $K\ge 2$,
$M\ge 3$, the separability criterion we can deduce is weaker, as PPT criterion is
no longer sufficient and necessary for the separability of bipartite states.
Nevertheless the canonical representation of PPT states can shade light on
studying the structure of bound entangles states which are PPT but
not separable.


\vskip 8mm
\begin{thebibliography}{20}

\bibitem{DiVincenzo}D.P. DiVincenzo,
{ Science} { 270}, 255 (1995).

\bibitem{teleport} C.H. Bennett, G. Brassard, C. Cr\'epeau,
       R. Jozsa, A. Peres
       and W.K. Wootters, { Phys. Rev. Lett.} { 70}, 1895
       (1993).\\
G.M. D'Ariano, P.Lo Presti, M.F. Sacchi, {\it Phys. Lett.} { A
272} (2000), 32.\\
S. Albeverio and S.M. Fei,
          { Phys. Lett.} { A 276}(2000)8-11.\\
S. Albeverio and S.M. Fei and W.L. Yang, { Commun. Theor.
Phys.} { 38} 301-304 (2002); { Phys. Rev.} { A 66} 012301
(2002).

\bibitem{densecrypto} C.H. Bennett and S.J. Wiesner,
        { Phys. Rev. Lett.} { 69}, 2881 (1992).\\
A.~Ekert, { Phys. Rev. Lett.} { 67}, 661 (1991).\\
D.~Deutsch, A.~Ekert, P. Rozsa, C. Macchiavello,
S. Popescu and A. Sanpera, { Phys. Rev. Lett.} { 77}, 2818
(1996).\\
C.A. Fuchs, N. Gisin,
         R.B. Griffiths, C-S. Niu, and
         A. Peres, { Phys. Rev. A}, { 56}, 1163 (1997).

\bibitem{Werner} R. Werner, Phys. Rev. A 40, 4277 (1989).

\bibitem{AB3H} G. Alber, T. Beth, M. Horodecki, P. Horodecki,
R. Horodecki, M. Rotteler, H. Weinfurter, R. Werner, and A.
Zeilinger, Quantum Information:An Introduction to Basic
Theoretical Concepts and Experiments(Springer,2001).

\bibitem{BCH} D. Bruss, J. I.Cirac, P. Horodecki, F. Hulpke,
B. Kraus, M.Lewenstein, and A. Sanpera, J. Mod. Opt. 49, 1399-1418 (2002).\\
D. Bruss, J. Math. Phys. 43, 4237 (2002).

\bibitem{10} R. F. Werner: Phys. Rev. A 40, 4277 (1989).\\
S. Popescu: Phys. Rev. Lett. 72, 797 (1994).\\
S. Popescu: Phys. Rev. Lett. 74, 2619 (1995).

\bibitem{Bell64} J.S. Bell, Physics (N.Y.) 1 (1964) 195.

\bibitem{Peres96a}
A. Peres, Phys. Rev. Lett. {\bf 76} 1413 (1996).\\
K. \.Zyczkowski and P. Horodecki, Phys. Rev. A {\bf 58},  883
(1998).

\bibitem{2hPRA99} M. Horodecki and P. Horodecki, Phys. Rev. A 59 (1999) 4206.

\bibitem{cag99} N.J. Cerf, C. Adami and R.M. Gingrich, Phys. Rev. A 60
(1999) 898.

\bibitem{nielson01} M.A. Nielsen and J. Kempe, Phys. Rev. Lett. 86 (2001)
5184.

\bibitem{3hPLA223} M. Horodecki, P. Horodecki and R.
Horodecki, Phys. Lett. A 223 (1996) 1.

\bibitem{ter00} B. Terhal, Phys. Lett. A 271 (2000) 319.\\
M. Lewenstein, B. Kraus, J.I. Cirac and P. Horodecki, Phys.
Rev. A 62 (2000) 052310.

\bibitem{dps} A.C. Doherty, P.A. Parrilo and F.M. Spedalieri, Phys. Rev.
Lett. 88 (2002) 187904.

\bibitem{ru02} O. Rudolph, {\it Further results on the cross
norm criterion for separability}, quant-ph/0202121.\\
K. Chen and L.A. Wu, Quant. Inf. Comput. 3 (2003) 193.

\bibitem{chenPLA02} K. Chen and L.A. Wu, Phys. Lett. A 306 (2002) 14.

\bibitem{grc}
S. Albeverio, K. Chen and S.M. Fei, {\it Generalized reduction
criterion for separability of quantum states}, to appear in Phys.
Rev. A (2003).

\bibitem{hlpre} P. Horodecki, M. Lewenstein, G. Vidal and I. Cirac, Phys.
Rev. A 62 (2000) 032310.

\bibitem{hlpre1}
P. Horodecki, J.A. Smolin, B.M. Terhal and A.V. Thapliyal,
Theor. Comp. Sci. 292 (2003) 589-596.\\
S. Albeverio, S.M. Fei and D. Goswami, { Phys. Lett.} { A},
91-96(2001).\\
S.M. Fei, X.H. Gao, X.H. Wang, Z.X. Wang and K. Wu, { Phys.
Lett. A} { 300} (2002) 559-566.\\
S.M. Fei, X.H. Gao, X.H. Wang, Z.X. Wang and K. Wu, {\sf Int. J.
Quant. Inform.}, {1}(2003) 1-13.

\bibitem{22n}
S. Karnas and  M. Lewenstein, Phys. Rev. A 64, 042313 (2001).\\
S.M. Fei, X.H. Gao, X.H. Wang, Z.X. Wang and K. Wu, Phys. Rev. A
68 022315 (2003).

\bibitem{222n}
S.M. Fei, X.H. Gao, X.H. Wang, Z.X. Wang and K. Wu, Commun. Theor.
Phys.40 (2003) 515-518.

\end{thebibliography}
\end{document}